\def\aa{{A\&A}}

\def\aj{{AJ}}

\def\apj{{ApJ}}

\def\mnras{{MNRAS}}
\def\nat{{Nature}}

\newcommand{\haemsigma}{$M_{\bullet}$--$\sigma_*$\ }
\newcommand{\haemsun}{M_{\odot }}
\newcommand{\haekms}{\, {\rm km \, s}^{-1} }

\newcommand{\haeyr}{\,{\rm yr}}
\newcommand{\haeMsol}{\,{M_\odot}}

\def\haega{\mathrel{\hbox{\rlap{\hbox{\lower4pt\hbox{$\sim$}}}\hbox{$>$}}}}
\def\haela{\mathrel{\hbox{\rlap{\hbox{\lower4pt\hbox{$\sim$}}}\hbox{$<$}}}}

\documentclass[cup5b]{caps}
\usepackage{graphicx}
\usepackage{amssymb}
\usepackage{ociwsymp1}
\HeadText{M. G. Haehnelt}

\begin{document}
\pagenumbering{arabic}
  
\author[]{MARTIN G. HAEHNELT\\Institute of Astronomy, Cambridge}

\chapter{Joint Formation of Supermassive \\ Black Holes and Galaxies}

\begin{abstract}
The tight correlation between black hole mass and velocity dispersion 
of galactic bulges is strong evidence that the formation of galaxies 
and supermassive black holes are closely linked. I review the
modeling of the joint formation of galaxies and their central 
supermassive black holes in the context of the hierarchical 
structure formation paradigm.   
\end{abstract}

\section{Supermassive Black Holes in Galactic Bulges} 

\subsection{ SMBHs ---  Commonplace Rather Than Rare Disease}

The suggestion that galaxies may harbor supermassive black holes (SMBHs)
at their centers dates back to the 1960s (Lynden-Bell 1969; 
see Rees 1984 for a review of the older literature). Initially black
holes  were suggested to explain the large efficiencies of
transforming matter into radiation 
necessary  to sustain  the large luminosities  of bright active
galactic nuclei (AGNs). For a long time the evidence for the 
presence of SMBHs was mainly due to 
observations of such ``active'' SMBHs.  However, the space density 
of bright AGNs is about a  factor of 100 to 1000 smaller than  that of 
normal galaxies. For several decades it was thus a lively
debated question whether all normal galaxies (including our own)
contain  ``dormant'' SMBHs, or whether only a small fraction  of 
all galaxies and which galaxies may exhibit such a ``disease''
(see Rees 1990 for a discussion). 

The discovery of the tight correlation between black hole mass and the 
velocity dispersion of the bulge component of galaxies (Gebhardt et al. 2000; 
Ferrarese \& Merritt 2000), enabled by the increased resolution
of {\it HST}, has decided 
this question beyond any doubt in favor of most  galactic 
{\it bulges}\  containing  black holes.  Note that this also finally 
rules out the possibility that bright QSOs are long-lived 
(e.g., Small \& Blandford 1992; Choi, Yang, \& Yi 1999).
In light of the faint-level AGN activity  that has been detected in 
many  galaxies for some time (see Ho, this volume, for a review),
this may not come too much as a surprise. It even appears 
now that not only most galaxies 
with a bulge contain SMBHs holes, but also that 
most of such galaxies show activity (see Heckman and Ho, this volume). 

The fact that the mass of the black hole appears to scale with the 
mass of the galactic bulge rather than the mass of the galaxy 
or its dark matter (DM) halo is likely to be a major new clue in understanding  
the formation of galactic bulges and central SMBHs.

\subsection{The \haemsigma Relation: An Undeciphered Clue?} 

The tight correlation of the black hole mass with global structural 
properties of the galaxy, such as the velocity dispersion of its bulge, 
came as a big surprise. The \haemsigma relation is generally taken as strong 
evidence that the growth of SMBHs and the formation of galaxies go hand in
hand. This is reflected in the fact that this idea has become the theme of 
this  conference. 

Many suggestions have been made as to what may determine the mass of SMBHs
in galaxies. The models can  be broadly grouped as follows.  

\begin{itemize}

\item Simple scaling models: The mass of the black holes is assumed 
      to scale with certain properties of the galaxy or its DM halo
      (e.g., Haehnelt \& Rees 1993; Haiman \& Loeb 1998). 

\item
Supply-driven models: The mass of accreted gas scales with  the amount of 
available (low-angular momentum) fuel (e.g., Kauffmann \& Haehnelt 2000; 
Adams, Graff,  \& Richstone 2001; Volonteri, Haardt, \& Madau 2002).

\item
Self-regulating models: The energy output during accretion or the influence of 
the black hole on the galactic gravitational potential limits the fuel supply 
(e.g., Norman \& Silk 1983; Small \& Blandford 1992; Silk \& Rees 1998; 
Haehnelt, Natarajan, \& Rees 1998; Sellwood \& Moore 1999; Wyithe \& Loeb 
2002; El-Zant et al. 2003).

\item 
``Exotic'' models: Accretion of collisional DM or accretion of stars (e.g.,
Hennawi \& Ostriker 2003; Zhao, Haehnelt, \& Rees 2002). 

\end{itemize} 

Most of these models succeed in explaining the observed scaling relations.  
Nevertheless, none appears likely to be the final answer.  Some of them fall 
short of specifying the physical mechanism at work.  The rest invoke 
fine-tuning or only moderately plausible assumptions of some kind or another. 
Most puzzling in that respect is the  remarkable tightness of the  observed 
correlation of the black hole mass with the stellar velocity dispersion on 
large radii. This correlation is significantly tighter than, for example, the 
correlation of stellar velocity dispersion and bulge luminosity (Haehnelt \& 
Kauffmann 2000).  It appears that either the tightness of the relation for the 
current sample of nearby galactic bulges is a statistical fluke or that we 
have not yet fully unravelled the clue it may give us as to how SMBHs 
grow and form. 

\begin{figure*}[t]
\includegraphics[width=1.00\columnwidth,angle=0]{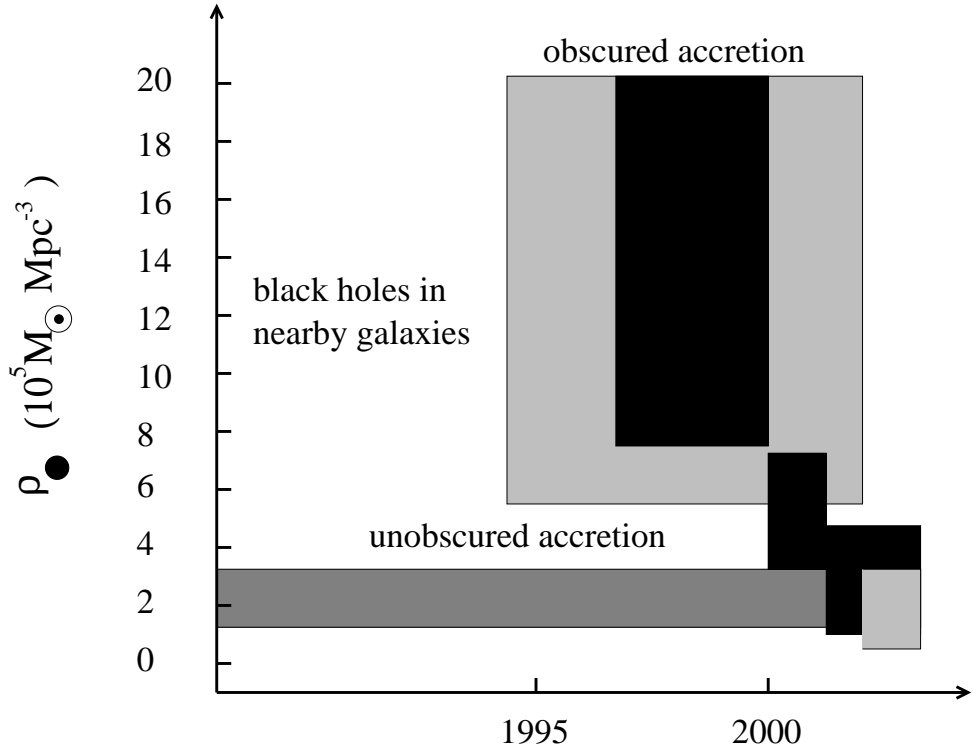}
\vskip 0pt \caption{
Comparison of estimates of black hole mass density 
for nearby galactic bulges (black) with that inferred for accretion 
in optically bright QSOs (unobscured accretion, dark grey) and 
in hard X-ray selected active SMBHs (obscured accretion, light grey).  
Estimates are converging to within a factor of 2 to the value 
inferred from optically bright QSOs.
\label{sample-figure}}
\end{figure*}

\section{The Assembly of SMBHs} 
\subsection{The Observed Accretion History of SMBHs} 

The next important piece of observational evidence on how SMBHs have assembled 
comes from the demography of the active population.  The observed
energy density $U_X$  emitted by active black holes at
redshift $z$ in band X  can be used to estimate  the 
overall black hole mass density in remnant black holes using So\l tan's (1982) 
argument

\begin{equation} 
\rho_{\bullet} = \epsilon_{\rm acc}^{-1} f_{X}^{-1}  \int{
\frac{dU_X(z)}{dz} (1+z) dz}.   
\end{equation} 

This elegant argument only requires assumptions about the efficiency of 
turning accreted rest-mass energy into radiated energy $\epsilon_{\rm acc}$ 
and the  factor $f_{X}$, which relates the total emissivity to the emissivity 
in the chosen band. For optically bright QSOs these estimates have been 
done for some time and the black hole mass density 
associated with accretion in  optically bright QSOs 
is well  determined. The remnant black hole mass 
density attributable to optically ``obscured'' accretion, 
as traced by hard X-ray sources, is much less well known. 
In the last couple of years it  has been claimed that the total 
mass accreted in optically  obscured  sources may exceed 
that accreted in optically bright  QSOs by a large factor 
(Fabian \& Iwasawa 1999; Elvis, Risaliti, \& Zamorani 2002). 
However, the bolometric correction factors were large ($\haega 30$) 
and uncertain, and the redshift
distribution was not yet determined. As discussed by Fabian (this volume), 
with the data on hard X-ray selected active SMBHs rapidly 
accumulating, a more moderate contribution to the integrated black
hole mass density appears more plausible. With this more moderate
contribution and the  reduced estimates of the black hole
mass density  in nearby galactic bulges that accompanied the 
discovery of the \haemsigma relation, it appears that  
optically bright QSOs are a reasonably faithful tracer of the
integrated mass accretion history of SMBHs. 
Figure~1.1 shows schematically how in the last few years estimates 
of the black hole mass density in nearby galactic bulges and in active
SMBHs have converged to within a factor of  2 to the value 
estimated for accretion in optically bright QSOs (Haehnelt \&
Kauffmann 2001; Merritt \& Ferrarese 2001;  Aller \& Richstone 2002; 
Yu \& Tremaine 2002; Comastri 2003; Fabian, this volume). 

\subsection{Theoretical Models for the Assembly of SMBHs} 

A wide variety of formation mechanisms for SMBHs
has been  suggested, which are nicely summarized in the famous flow 
chart of Martin Rees (Rees 1977, 1978, 1984). The suggestions 
vary from the collapse of a (rotating) supermassive stars, to 
the evolution of a dense star cluster,  to the direct collapse of a 
gas cloud, and various combinations thereof. The main message of the 
flow chart, that there is no lack of plausible routes to a SMBH,
is as valid today as it was when it was first compiled.  
However, for most of the processes the large formation efficiency of 
$\sim 0.2$\% of the baryonic mass in a galactic bulge and the tight
correlation with the stellar velocity dispersion of the galactic 
bulge are surprising. {\it Have we, then, made progress and can  we 
exclude large parts of the diagram?} Probably not much. What has been 
established so far is  that the majority of  galactic bulges 
contain SMBHs and that emission of optically bright QSOs 
trace the mass accretion history well.  We further 
know from our theoretical understanding of the hierarchical
build-up of galaxies that  mergers of SMBHs 
have to play an important  role in their build-up. However, at the same 
time, bright, high-redshift QSOs tell us that about 10\%
of the black hole mass density in SMBHs had already been assembled 
at redshift 4, and some black black holes already had a mass
of $10^9 \haeMsol$ even at redshift 6. 

Thus, a consistent picture for the  growth history of SMBHs 
at redshift $z\haela 5$  appears to be in place. 
The more difficult  and also more interesting  problem of 
how SMBHs  have formed  in the first place is, however,  still largely
unconstrained by observations. 

\begin{figure*}[t]
\includegraphics[width=0.95\columnwidth,angle=0]{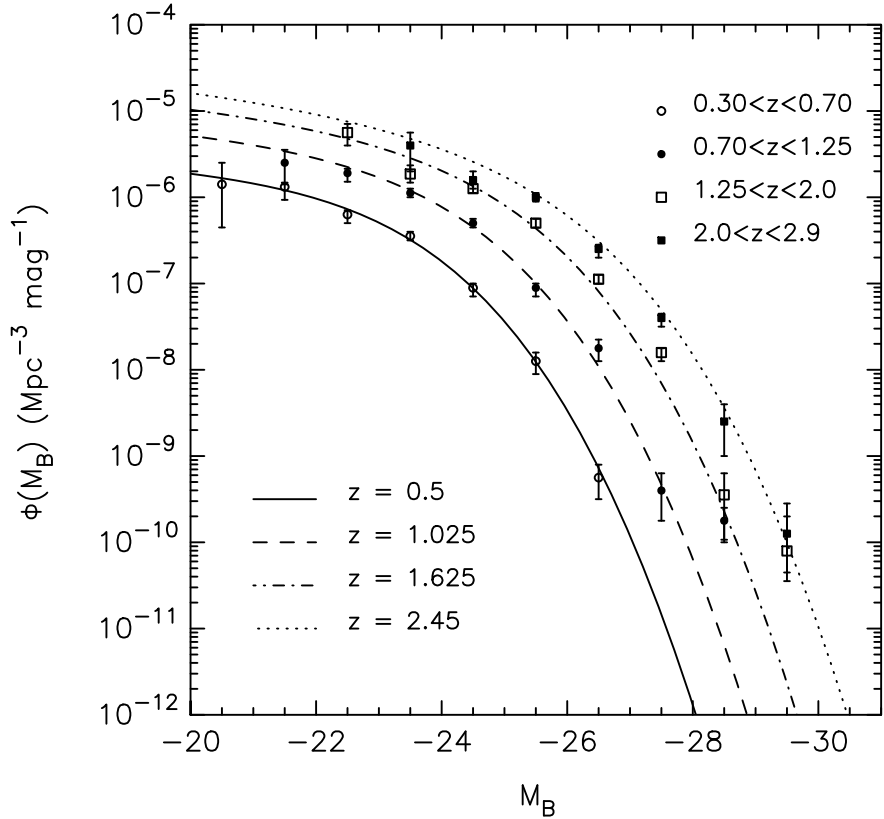}
\vskip 0pt \caption{
The luminosity function of QSOs in a model where the 
space density of AGNs is related to the formation rate of DM 
halos . (Reproduced from Haehnelt \& Rees 1993.) 
\label{sample-figure}}
\end{figure*}

\section{Hierarchical Galaxy Formation}
\subsection{The Standard Paradigm of Structure Formation} 

The joint theoretical and observational effort 
of the last three decades has led to a determination or, maybe better,
reconstruction of the spatial power spectrum of the initial 
density fluctuations, which is reliable to better than 20\% on 
scales from a few Mpc to the Hubble radius (see Spergel et al. 2003 
for a recent summary).  This power spectrum
of density fluctuations is  hierarchical in nature: fluctuations
on large scales are smaller  than those on small scales. The power
spectrum is  tantalizingly close to the $\Lambda$CDM variant of 
the cold dark matter (CDM) scenario, which postulates DM particles 
with random velocities  too small to affect structure formation. Such 
a power spectrum results in a 
hierarchical  build-up of structures with small objects forming
first and larger objects building  up by merging.  In this model the 
first pregalactic structures form at redshift 20 to 5, while the first 
DM halos capable of hosting big galaxies assemble between redshift 5 
and 2,  and the formation of galaxy clusters occurs at redshift 1 to 0.

\subsection{Hierarchical Build-up of Galaxies and Merger Tree Models}

The hierarchical structure formation paradigm has profound
implications for the way galaxies and active nuclei form and evolve
(White \& Rees 1978; Blumenthal et al. 1984; White \& Frenk 1991; 
Haehnelt \& Rees 1993).  Galaxies typically double their mass in 
about 20\% of the Hubble time. 
A key ingredient of hierarchical galaxy formation models is thus 
the frequent merging of DM halos and galaxies. 
Kauffmann and collaborators were the 
first to model in detail the basic processes governing galaxy formation 
following the merging history  
of  DM halos  (Kauffmann, White, \& Guiderdoni 1993). Gas cooling, star 
formation, and stellar feedback were 
described with simple recipes, and the merger history was modeled with 
extensive Monte Carlo simulations. This model soon became known as the 
``semi-analytical model'' of galaxy formation. It was shown 
to be consistent with a wealth of observations at low redshift 
if the parameterization of the physical processes are 
chosen in an appropriate way
(e.g., Cole et al. 1994;  Kauffmann et al. 1999a; 
Somerville \& Primack 1999; Somerville, this volume). 
High-redshift observations have 
removed  some, but not all, of the parameter  degeneracies 
that exist (Kauffmann et al. 1999b; 
Somerville, Primack, \& Faber 2001). The detailed 
comparison with the luminosity function and clustering 
properties of galaxies in different bands nevertheless gives 
confidence that the models are a fair representation 
of the hierarchical build-up of the {\it mass} of  
observed galaxies, their bulges, and their DM halos. Details 
of the predicted star formation history, luminosities, and colors at 
high redshift  are, however,  much less certain. 

Despite their success in reproducing many observed properties of
galaxies, the fact that stellar populations in early-type galaxies 
are generally very old regardless of their location in galaxy 
clusters or in the field 
(e.g., van~Dokkum et al. 2001) has not yet been fully understood in the
context of hierarchical galaxy formation models. I will discuss 
later how  this may be related to a similar problem of understanding 
the  rapid decline of the QSO emissivity at low redshift. 

%\begin{figure*}[t]
%\includegraphics[width=0.95\columnwidth,angle=0]{haehneltc02_fig3.ps}
%\vskip 0pt \caption{
%``Merger tree' 'models of galaxies with black holes: 
%Monte Carlo realization of the merging histories of DM
%halos plus simple recipes for gas cooling, star formation, 
%stellar feedback accretion, and dynamics of galaxies and black holes.
%\label{sample-figure}}
%\end{figure*}

\section{SMBHs in Hierarchically Merging Galaxies} 

\subsection{SMBHs in DM Halos} 

The merging histories and dynamical evolution of DM halos in CDM-like 
structure formation models has been  well established by analytical 
calculations and numerical simulations (e.g., 
Jenkins et al. 1998). As mentioned in the last
section, the evolution of the baryonic component of these DM halos is 
more difficult to predict. It is thus reasonable to try to  link
the evolution of the population of (active) SMBHs directly to the
evolution of the population of DM halos, especially  at high
redshift where our knowledge about the host galaxies of SMBHs  
has only recently started to build up.  When CDM-like  models were first
seriously discussed, it was not obvious that the rather late emergence
of galaxy-size structures in these models was not in conflict with the 
existence of bright, high-redshift QSOs, which were found in large 
numbers at the end of the 1980s.  Efstathiou \& Rees (1988) showed that
this is not the case. Haehnelt \& Rees (1993) pointed out that 
in CDM models the build up of galactic-size structures coincides 
with the peak in AGN activity at redshift 2.5 and that 
the emergence of the QSO population  
between redshift 5 and 2.5 arises naturally in CDM-like
cosmogonies (see also Cavaliere \& Szalay 1986). 
They further  showed that the evolution of the QSO  
emissivity  at $0<z<3$ can be reproduced if a 
suitable scaling of black
hole mass with the virial velocity of newly formed halos and redshift 
was assumed (Fig. 1.2). The formation  rate of galactic nuclei 
was modeled as the time derivative of the space density of 
DM halos times a  ``lifetime''  $t_{\rm Q}$ of the QSO phase,  
while the luminosity was assumed to scale with the Eddington luminosity: 
\begin{eqnarray} 
N_{\rm QSO} &=& \dot N_{\rm halo}  t_{\rm Q},  \nonumber\\
L_{\rm QSO} &=& \alpha_{\rm Edd} L_{\rm Edd}(M_{\bullet}),\\ 
M_{\bullet} &=& g(M_{\rm halo},z) = g({\upsilon}_{\rm circ},z) \nonumber 
\end{eqnarray} 
Haehnelt \& Rees (1993) had to choose the  scaling with redshift such 
that the black hole formation efficiency  dropped very quickly with 
decreasing redshift. They argued that this is because at redshifts
smaller than 2.5 the amount of available fuel in galaxy-size halos  
decreases with the decreasing 
ability of the gas to cool at low redshift 
(see also Cavaliere \& Vittorini 2000 and see 
Monaco, Salucci, \& Danese 2000 for an alternative interpretation). 
They further found that $t_{\rm Q} \approx 
10^{8}  \haeyr$ gives a good match to the 
QSO luminosity function for their assumed  scaling of black hole formation
efficiency with virial velocity of the DM halo. 
In a similar spirit, Haiman \& Loeb (1998) suggested  
that the black hole mass scales linearly with halo mass 
and showed that the QSO luminosity function can then be 
reproduced with a very short lifetime $t_{\rm Q} \approx 10^{6} \haeyr$. 
Haehnelt et al. (1998) showed that a consistent  picture 
for the high-redshift star-forming (``Lyman-break'') galaxies
and the AGN population arises within this  framework if 
Lyman-break galaxies host the  SMBHs responsible for the
observed QSO activity (see Granato et al. 2001 for a discussion 
of the possible connection between QSOs and sub-mm sources).  
Haehnelt et al. (1998) showed that the scaling 
of  black hole mass with halo mass or circular velocity is 
degenerate with the assumed lifetime $t_{\rm Q}$. 
The QSO luminosity function can be matched either with a steep relation
between black hole mass and halo mass ($M_{\bullet} \propto M_{\rm
halo}^{5/3} \propto {\upsilon}_{\rm circ}^{5}) $, more massive halos  for a
given black hole mass and a longer lifetime {\it or} a shallower 
relation between 
black hole and halo mass  ($M_{\bullet} \propto M_{\rm halo}\propto 
{\upsilon}_{\rm circ}^{3}$), less massive halos and a
shorter lifetime. Haehnelt et al. (1998) further showed  that this degeneracy 
can be broken with a study of the clustering properties of AGNs, which
is sensitive to the absolute mass scale of the DM halos hosting
QSOs. They  found  that a scaling  $M_{\bullet} \propto {\upsilon}_{\rm circ}^{5}$ 
and  $t_{\rm Q} \approx 10^{7} \haeyr$ did fit best the QSO luminosity function 
and the then still rather scarce  information on the clustering of 
high-redshift QSOs.  Haiman \& Hui (2001)  and Martini \& 
Weinberg (2001) 
investigated in more detail what constraints on the lifetime can 
be obtained from upcoming QSO surveys (see Martini, this volume,  for a 
summary of this and other constraints on QSO lifetimes). 
Wyithe \& Loeb (2002)  showed that a scaling 
$M_{\bullet}  \propto {\upsilon}_{\rm circ}^{5} $ is also 
consistent with the new high-redshift data from the SDSS 
QSO survey and the luminosity function of X-ray selected AGNs. 

\begin{figure*}[t]
\includegraphics[width=0.95\columnwidth,angle=0]{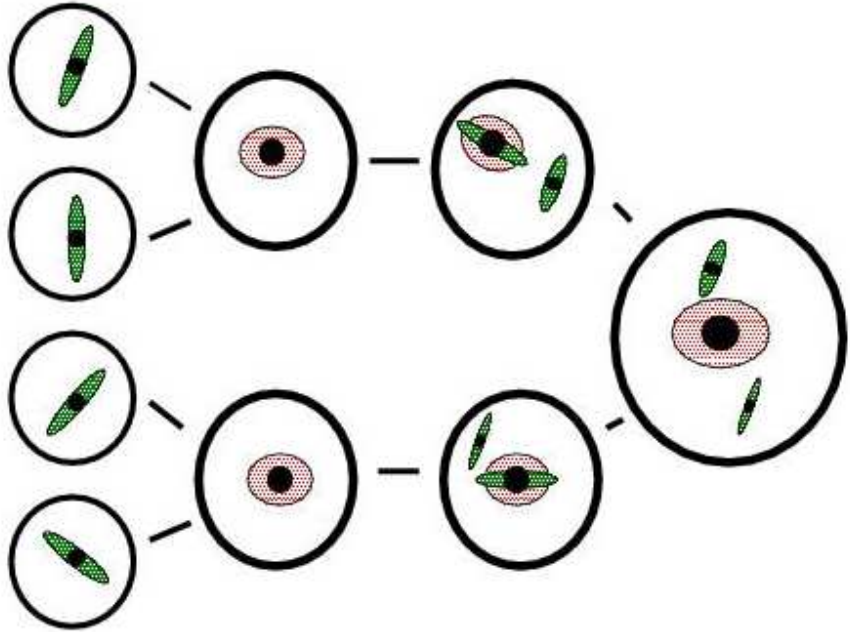}
\vskip 0pt \caption{
``Merger tree' 'models of galaxies with black holes:
Monte Carlo realization of the merging histories of DM
halos plus simple recipes for gas cooling, star formation,
stellar feedback accretion, and dynamics of galaxies and black holes.
\label{sample-figure}}
\end{figure*}

\begin{figure*}[t]
\includegraphics[width=1.00\columnwidth,angle=0]{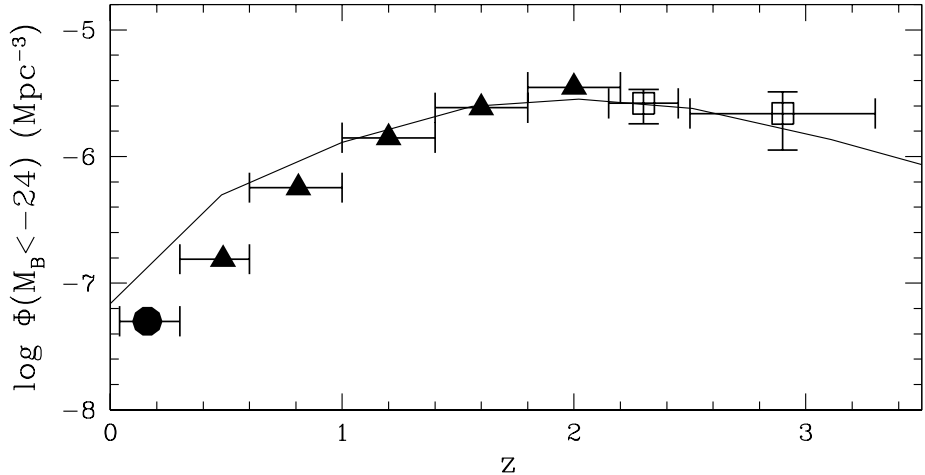}
\vskip 0pt \caption{
Redshift evolution of the space density of QSOs with $M_{\rm B}
<-24$ mag in the model of Kauffmann \& Haehnelt (2000).
(Reproduced from Haehnelt \& Kauffmann 2001.)
\label{sample-figure}}
\end{figure*}

\subsection{SMBHs in Merger Tree Models}

When  Magorrian et al.  (1998) published a  large sample 
of black hole mass estimates  for nearby galaxies and confirmed 
the suggestion that  black hole mass and bulge mass are strongly 
correlated with a nearly linear relation (Kormendy \& Richstone 1995), 
the idea that the formation of galaxies and black holes are closely 
linked became more widely accepted. Cattaneo, Haehnelt, \& Rees  (1999) 
demonstrated that a  hierarchical merger tree model could reproduce 
the observed slope and scatter of the ``Magorrian relation,'' provided that 
star  formation and the growth of black holes are closely linked.  
They thereby  enforced a rapid drop of the   gas available in both shallow 
and very deep potential wells to mimic the effect of feedback on 
small halos and the inability of the gas to cool in 
the very massive halos. 
Kauffmann \& Haehnelt (2000) combined the  full merger tree models 
of galaxies with the idea that the SMBHs form from  the
cold gas available in major mergers and assumed  that the accretion 
of gas leads to QSO activity of duration $t_{\rm Q} \approx 10^{7} \haeyr$. 
They demonstrated that such a 
model is  consistent with the evolution of the QSO emissivity
(Fig 1.4.), the Magorrian relation, the present-day luminosity function, 
and global star formation history of galaxies and the evolution of cold gas as 
probed by damped Ly$\alpha$ systems. It is interesting to note that to match 
the observations Kauffmann \& Haehnelt (2000) had to change the feedback
prescription compared to the previous modeling of Kauffmann et al.,  
such that galaxies become progressively more gas rich at high redshift. 
Kauffmann \& Haehnelt (2000) also made some predictions 
for  the luminosities of QSO host galaxies at high redshift. 
In the model of Kauffmann \& Haehnelt the rather steep decline 
of the QSO emissivity with decreasing redshift can be 
attributed to a combination of  a decrease in the merger rate, 
a decrease of  the amount  of cold gas available for fueling,  
and an increase in the accretion time scale.  

\begin{figure*}[t]
\includegraphics[width=1.00\columnwidth,angle=0]{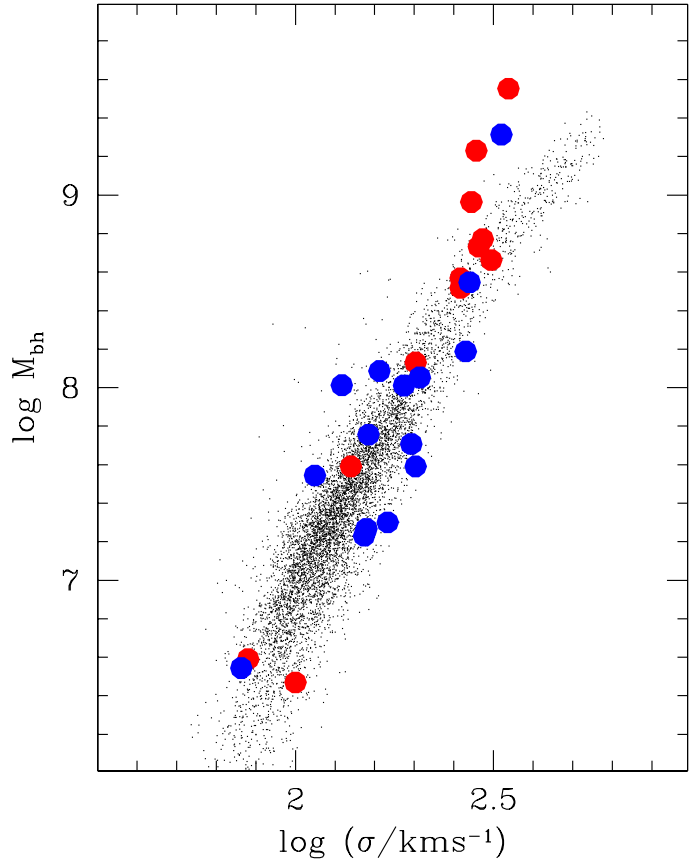}
\vskip 0pt \caption{
The \haemsigma relation in the model of Kauffmann \& Haehnelt (2000).
(Reproduced from Haehnelt 2002.)
\label{sample-figure}}
\end{figure*}

Haehnelt \& Kauffmann (2000) showed that the Kauffmann \& Haehnelt (2000) 
model is also 
consistent with the \haemsigma relation (Fig. 1.5). Somewhat surprisingly, 
the model also reproduces the tightness of the 
relation, despite the frequent merging. This is because the black holes 
move along the \haemsigma relation during mergers and the cold gas
available for accretion during major mergers scales explicitly 
with the  depth of potential well rather than with the mass of the
galaxy in their model. One should, however,  keep in mind that the 
dynamical processes were modeled in a simplistic fashion. More 
detailed modeling  
of the dynamics of the galaxy and black hole merger 
most likely will introduce additional scatter. 
Kauffmann \& Haehnelt (2002) investigated the clustering properties 
of galaxies and QSOs in the model and found that a lifetime of 
$10^{7}$ yr is also consistent with the 2dF clustering data. They further 
pointed out that study of  the galaxy-QSO correlation  function 
may give further clues on  how QSO activity is triggered. 
The spin distribution of black holes may give another handle 
on determining the merger history of black holes (Hughes \& Blandford
2003). Cattaneo (2002) investigated the expected evolution of the spin
of black  holes hosted by hierarchically merging DM halos for a variety of 
assumptions  for the spin-up and spin-down processes. Unfortunately,  
observationally we know very little about the spin of SMBHs
even though it has been suggested that the jets in radio-loud 
AGNs may be powered by the rotational energy of their central black
hole (Blandford \& Znajek 1977).

Despite their success in reproducing many observed properties, 
hierarchical models generally struggle to reproduce the rapid decline 
of QSO activity
toward low redshift. As mentioned earlier, this difficulty is most likely  
connected to a  similar difficulty in explaining the  
very rapid decrease of star formation activity in galactic 
bulges. In galactic bulges both the star 
formation activity and the accretion onto the black holes 
seem to be ``switched off'' at higher redshift 
than star formation in other galaxies.  This suggests 
that the presence of a supermassive black hole is the 
physical reason (see Granato et al. 2001 for a model along 
those lines). Feedback effects like radiation pressure, 
radiative heating, energetic  particles, and kinetic energy
due to accretion  have a large impact 
on the surrounding (forming) host galaxy and beyond
(see Begelman, this volume, for 
a review).  A better understanding  of AGN feedback may thus  be 
key to resolving  some of the difficulties encountered by 
the current models (see also Somerville, this volume).  
Supermassive black holes may so be  an essential ingredient 
in shaping the Hubble sequence of galaxies. 

\subsection{Binary/Multiple SMBHs and the Core Properties of
Galactic Bulges} 

If galaxies merge hierarchically and all galactic bulges contain black
holes the formation of supermassive binary and multiple black holes 
is expected to be common.  When two galaxies of moderate mass
ratio merge,  a hard binary will form quickly (e.g., Miloslavjevi\'c 
\& Merritt  2001). There are two processes that can shrink the
separation of hard binaries:  accretion of gas and hardening by 
three-body interaction with stars (Begelman, Blandford, \& Rees 1980). 
Stellar hardening of supermassive binaries requires the ejection of 
several times the mass of the binary black hole. Binary black 
holes may thus play an
important role in shaping the core profiles of galactic bulges 
(Ravindranath, Ho, \& Filippenko 2002; Milosavljevi\'c et al. 2002; 
Merritt, this volume).  Disk accretion of cold gas 
following minor mergers has been argued to be an efficient mechanism
to merge  binary black holes with a small ratio of secondary to primary 
mass (Armitage \& Natarajan 2002). The formation and evolution of 
a supermassive binary black hole in a  major merger of gas rich 
galaxies has not yet been studied, but it seems likely that gas
accretion will also lead to rapid hardening in this case. 

Whether binary black holes in typical low-redshift galaxies 
can reach the  separation at which emission of 
gravitational radiation leads to coalescence within a Hubble 
time is somewhat uncertain (Yu \& Tremaine 2002),  but it appears 
likely that they do in all but the most massive  elliptical galaxies 
(Haehnelt \& Kauffmann 2002; Milosavljevi\'c \& Merritt 2003).  
Observationally hard  binary  black holes are difficult to detect, and 
observational evidence  so far is circumstantial  
(Merritt \& Ekers 2002; see Komossa 2003 for a review).  

In hierarchically merging galaxies there is a significant 
probability that a third black hole will fall in before a hard binary 
black hole has coalesced. This  will normally  lead to 
gravitational slingshot ejection of the lightest black hole (Saslaw, Valtonen, 
\& Aarseth 1974; Hut \& Rees 1992). The binary will also  get a 
kick velocity. If all three black holes have similar masses,  
the kick velocity will be sufficient to kick the binary into the 
outer parts of the galaxy or even to eject it entirely 
(Hut \& Rees 1992; Xu \& Ostriker 1994).  As discussed by 
Redmount \& Rees (1989), at  coalescence supermassive binaries 
will also get a kick due to the radiation-reaction forces  
predicted by general relativity. The kick velocity due to the
resulting recoil should scale linearly with the mass ratio of 
the coalescing binary for small mass ratios. 
The absolute values of the kick velocities are uncertain. They  will 
depend strongly  on the radius of the last stable orbit and therefore 
on the spin and orbital orientation of the binary (Fitchett \&
Detweiler 1984).  While the expected value for a Schwarzschild
hole is small for rapidly spinning black holes, the kick velocity may be
as large  as a few  hundred $\haekms$ and may  remove  the  merged
binary from small and maybe even big galaxies in some cases. Note
that in the case of  two spinning black  holes  
the asymmetry of the radiation with respect to 
the plane of the orbit may result in a recoil along the direction of 
the orbital angular momentum (Redmount \& Rees 1989). Three-body 
interaction and the gravitational-radiation recoil of SMBHs 
in hierarchically merging galaxies may thus lead 
to a population of black holes outside galaxies 
(see also Haehnelt \& Kauffmann 2002; Volonteri et al. 2002).

\begin{figure*}[t]
\includegraphics[width=1.00\columnwidth,angle=0]{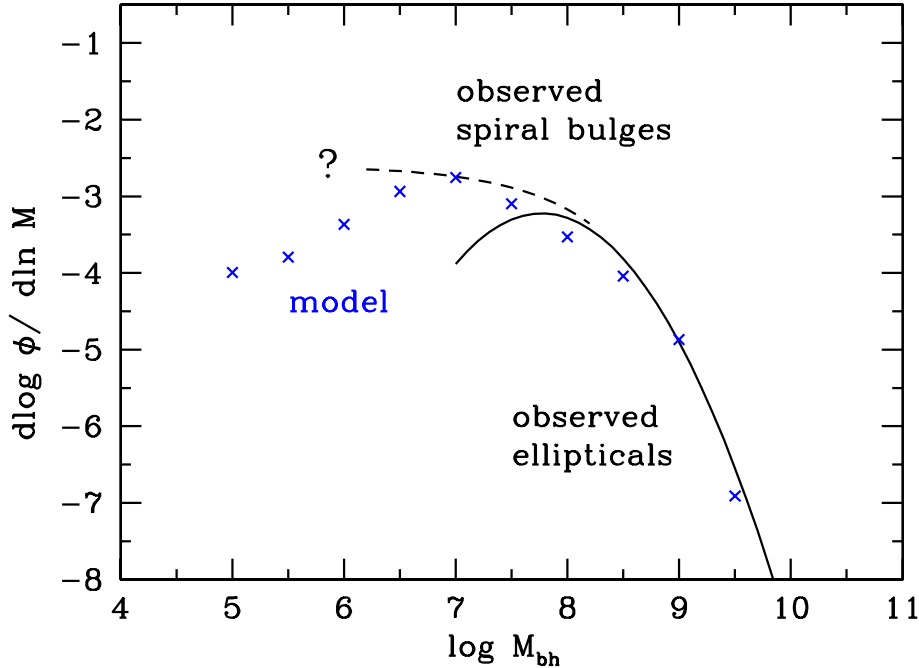}
\vskip 0pt \caption{
The mass function of the black holes in the model
of Haehnelt \& Kauffmann (2000) compared to the mass function inferred for
early-type galaxies extrapolated to include the contribution from
spiral bulges. (Adopted from Haehnelt 2003.)
\label{sample-figure}}
\end{figure*}

\begin{figure*}[t] 
\includegraphics[width=1.00\columnwidth,angle=0]{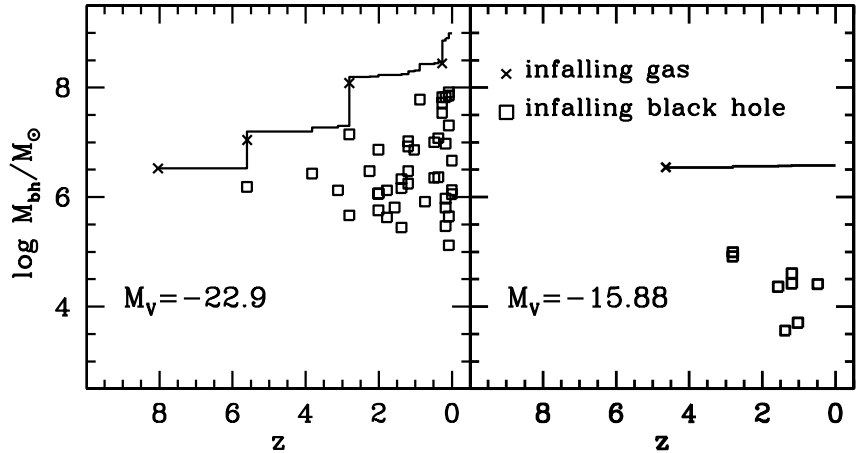}
\vskip 0pt \caption{ 
Typical assembly history of a SMBH for a bright and a faint bulge in the
model of Haehnelt \& Kauffmann (2000).  (Reproduced from
Haehnelt 2003.)
\label{sample-figure}}
\end{figure*} 

\section{Early Evolution}

\subsection{Do Intermediate-mass Black Holes Form in Shallow Potential 
Wells?} 

While the accretion history of SMBHs between
redshift 5 and 0 seems now reasonably well constrained, 
little is known about the the accretion/formation history  at higher  
redshifts.  At redshift 5 the mass of a typical DM halo in the 
$\Lambda$CDM  model is about $10^{11} \haemsun$. In the  $\Lambda$CDM 
model the matter fluctuation spectrum extends to smaller masses, and 
hierarchical growth of structure starts at much earlier times. The 
recent detection of a large polarization signal in the CMB at large 
scales by the {\it WMAP}\ satellite  requires that a large volume fraction of 
the Universe was reionized at $z\approx 20$ (Kogut et al. 2003); however,
note that the measurement error is still large. 
If confirmed, this is strong observational evidence that 
the matter fluctuation power  spectrum extends to scales 
as small as $10^{6}\, \haemsun$, or even
smaller. On these  scales nonlinear  structures form   at $z>20$.  
{\it Should we expect 
hierarchical growth at similarly early times to contribute
significantly to the build-up of SMBHs?}  
This depends crucially on  what happens in shallow potential wells 
with circular velocities ${\upsilon}_c \haela 100 \, \haekms$.  Modeling of 
the galaxy luminosity function in a CDM-like hierarchical structure formation
model requires that the star formation efficiency drops rapidly in 
shallow potential wells. Otherwise the  
faint-end slope of the luminosity function would be much steeper than 
observed (White \& Frenk 1991). The effect of stellar feedback and/or 
the effect of  heating due to photoionization is  generally  invoked  
as explanation. It appears plausible that SMBHs  assemble from the same 
low-angular momentum  tail of the cold gas  reservoir that also fuels 
star formation. This was, for instance, assumed  in the model of Kauffmann \& 
Haehnelt (2000), in which  SMBHs only form
efficiently in halos with  ${\upsilon}_c \haega 100 \, \haekms$. As a result,  
the mass function of black holes turns over at masses 
$\haela 10^{7}\, \haeMsol$  (Fig. 1.6), and  the hierarchical build-up of 
black holes only starts  at 
redshift $z\haela 6-8$ in their model (Fig 1.7).  
Nevertheless, some star formation has to occur in more 
shallow potential if reionization starts as early as implied by 
the large electron optical depth found by {\it WMAP}.  
Madau \& Rees (2001) suggested that the black hole remnants of
massive stars formed in these shallow potential wells will lead to the 
formation of a population of intermediate-mass black holes.   
Volonteri et al.  (2002, 2003)  followed the merging of DM
halos  containing SMBHs from $z> 20$. 
They found that hierarchical build-up starting from stellar-mass 
seed black holes  can contribute  significantly to the  population of 
SMBHs at redshift five  {\it if} the black holes 
merge efficiently (see also Islam, Taylor, \& Silk 2003). 
Whether black holes can merge in these shallow potential  is, 
however,  very uncertain. It requires rapid sinking of
the seed black holes to the center of merged pregalactic structures 
{\it plus} either  a dense stellar system to provide a sufficient 
number of stars for binary hardening or the accretion  of cold gas.
Stars and the cold gas from which they form are, however, expected to
be in short supply in shallow potential wells. 
Furthermore, even if the  gravitational radiation recoil were too small
to affect  the hierarchical build-up of SMBHs in normal galaxies, it 
may nevertheless efficiently remove  a significant fraction of
coalescing  binary black holes from pregalactic structures 
with shallow potential wells, should they exist. 
Currently our best 
bet to make further progress in this area is to  study how the 
\haemsigma relation extends to smaller galaxies. 
There are a number of observational claims for the detection 
of intermediate-mass black holes  (King et al. 2001; Gebhardt, Rich, \& 
Ho 2002; Filippenko \& Ho 2003; van~der~Marel, this volume). 
Should these detections  consolidate, the next step will 
be to  establish whether intermediate-mass black holes  have formed 
with similarly large efficiency as the SMBHs in 
deep potential wells.  This would  argue for a continuous
hierarchical build-up from stellar-mass seeds at very high redshift. 
With  {\it JWST} it should be possible to probe some of this directly
(see Haiman, this volume).

\begin{figure*}[t]
\includegraphics[width=1.00\columnwidth,angle=0]{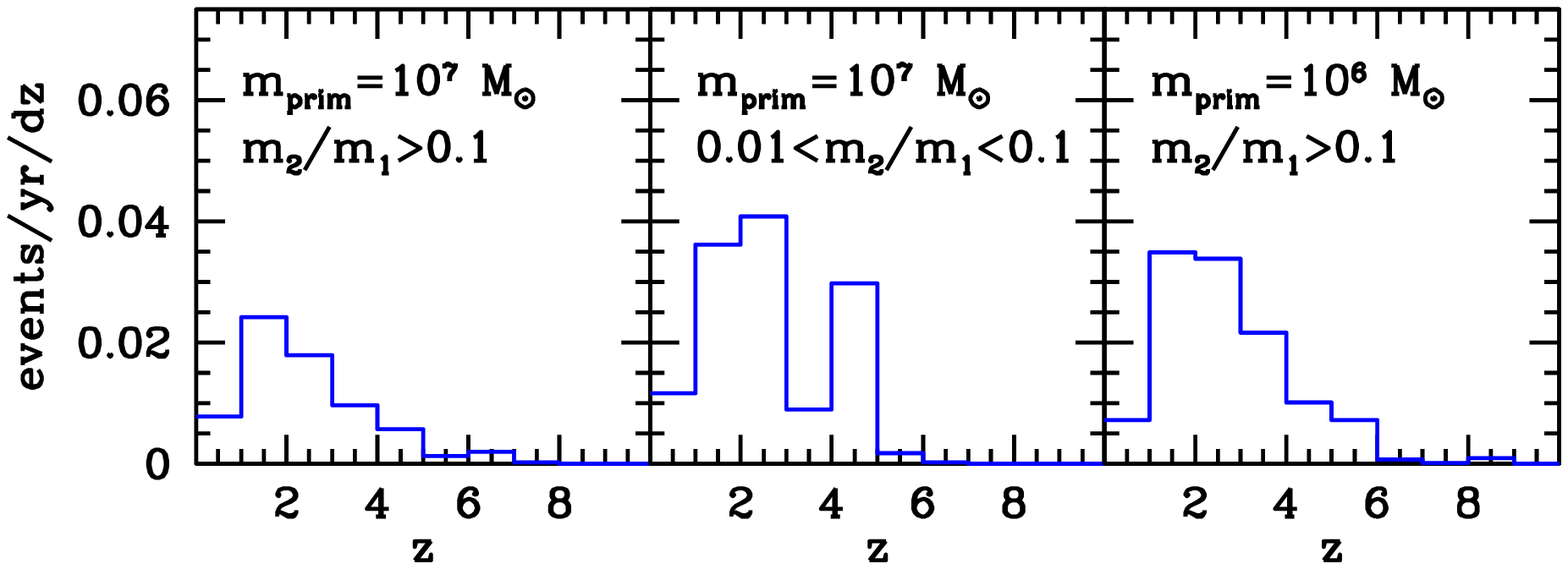}
\vspace{0.25cm}
\includegraphics[width=1.00\columnwidth,angle=0]{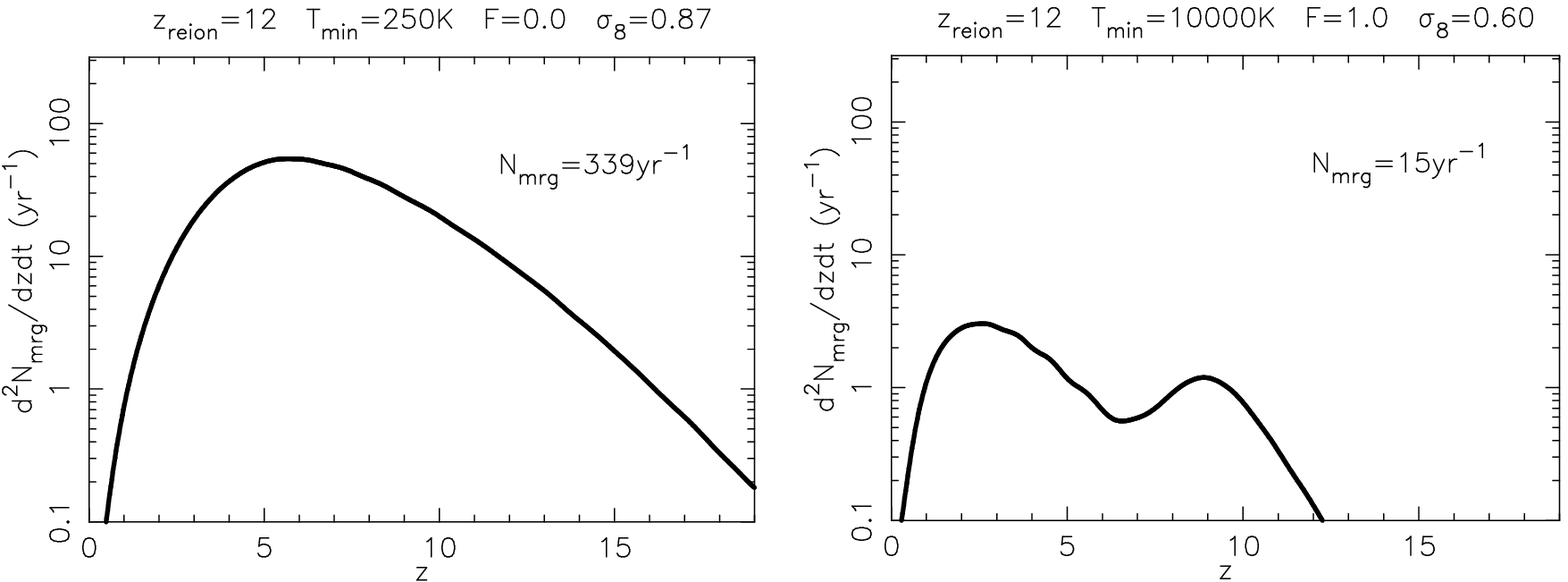}
\vskip 0pt \caption{
{\it Top:} Merging rate of galaxies 
forming supermassive binary black holes
for a range of primary black hole mass and mass ratio in the models of
Kauffmann \& Haehnelt (2000). (Reproduced from Haehnelt 2003.)
{\it Bottom:} Merging rate of DM halos hosting SMBHs predicted 
by  Wyithe \& Loeb (2002).
\label{sample-figure}}
\end{figure*}

\subsection{{\it LISA} and the Assembly of SMBHs at $z>5$} 

A somewhat longer shot to establish how the SMBHs 
observed at redshifts $\haega 5$  did assemble will be observations with the 
space-based gravitational wave interferometer {\it LISA}, expected to be 
launched in the next decade. {\it LISA}\ will be sensitive to gravitational 
waves with frequencies below 1 Hz, which are  not accessible from the ground 
because of seismic noise. The 
typical dynamical time at the  Schwarzschild radius of a 
SMBHs is $t_{\rm dyn} \sim 3 (M/10^{5}\,\haemsun)$ s.  The  detection of black 
hole-black hole mergers  involving SMBHs is thus a prime
objective of {\it LISA}.  The long baseline achievable in space means 
that for SMBHs {\it LISA}\ will not be so much   sensitivity limited 
but rather event-rate limited instead.  {\it LISA}\ will be sensitive 
enough to detect  equal-mass mergers of $10^4 - 10^7 \haeMsol $ black holes  
at $z< 20$. {\it LISA}\ will even be able to determine the luminosity distance 
and spin  for some coalescing binary black holes (Bender 2003).
However, if the black hole formation efficiency 
drops rapidly in potential wells with ${\upsilon}_{\rm circ}  <  100 \,\haekms $, 
event rates are only  
$0.3 -1 \haeyr^{-1}$ (Fig. 1.8). If hierarchical build-up extends  to 
smaller DM halos, event rates could be as large as  
few tens to a  hundred per year (Haehnelt 2003; see also Haehnelt
1994, 1998;  Menou, Haiman, \& Narayanan 2001; Wyithe \& Loeb 2002). 
Note, however, that this would predict a  mass function 
that rapidly rises toward smaller masses (Fig. 1.6). 
{\it LISA}\ has the exiting prospect to finally settle the question of 
how the SMBHs that power high-redshift AGNs were assembled.

\section{Open Questions} 

The recent observational progress has led to big improvements 
in our understanding on how black holes were assembled in 
hierarchically merging galaxies. There is, nevertheless, a long 
list of unanswered questions. The following is a certainly incomplete 
version of such a list:  

\begin{enumerate} 

\item
Is AGN activity really triggered by mergers (beginning, end, 
multiple)? What is the time scale of QSO activity? What determines 
it? Why is it apparently shorter than the merging time scale of galaxies? 

\item
How much room is there for dark or obscured  accretion? Can the accretion
rate exceed the Eddington limit?

\item
What is the physical origin of the \haemsigma relation? Is it as 
tight as claimed, and if so,  why?  Does it evolve with  redshift? 

\item
Does AGN activity affect the cooling/heating budget during galaxy 
formation in a global sense? What role do SMBHs play in
defining the Hubble sequence of galaxies?  

\item
Are (hard) supermassive  binary black holes common? On which time scale  do 
they merge? Are supermassive binary black holes  responsible for the core
properties of galactic bulges? Do black holes receive kick velocities 
that eject them  from (small) galaxies?

\item
Do intermediate-mass  black holes form in shallow potential wells? 
Does the \haemsigma relation extend to smaller black hole masses? 
Does the hierarchical build-up of SMBHs extend to pregalactic
structures at very high redshift?  
\end{enumerate}

Many of these questions are already under intense scrutiny. 
Progress in answering  them  will hopefully bring us closer to 
a more complete understanding of the physical processes responsible  
for the formation of SMBHs. 

\section{Summary} 

Models where black holes grow by a combination of gas accretion 
traced by short-lived ($\sim 10^{7} \haeyr$) QSO activity and
merging in hierarchically merging galaxies are consistent with a wide 
range of observations in the redshift range $0<z<5$. The rapid decline 
of both the QSO activity and star formation in galactic bulges 
suggests that SMBHs may play a bigger role in shaping the Hubble
sequence of galaxies than previously anticipated. 
The frequent merging of galaxies will lead to the ubiquitous
formation of supermassive binary black holes. These are generally 
expected to merge in all but the most massive galaxies by 
gas infall and stellar hardening. Gravitational slingshot 
in triple black holes and gravitational radiation recoil 
of coalescing binaries may lead to a substantial population 
of SMBHs outside of galaxies. The actual formation of SMBHs is still 
poorly constrained observationally. Direct collapse of a gas cloud to 
seed black holes  of $10^5-10^6 \haeMsol$ and extension of the
hierarchical merging all the way to stellar-mass seed black holes 
from Population III stars mark two extreme ends  of the list of viable 
possibilities. Further observational and theoretical study of  
intermediate-mass black holes, and eventually 
the study of the merging history of black holes with {\it JWST}\ and  the 
planned space-based gravitational wave interferometer {\it LISA}\
should help to  answer the question of how SMBHs formed.

\vspace{0.3cm}
{\bf Acknowledgements}.
I would like to thank Luis Ho for his patient 
persistence in urging me to write this review. Thank is also 
due to Stuart Wyithe for providing Figure 8{\it b}.

\end{document}